\documentclass[twocolumn,floatfix,tightenlines,superscriptaddress]{revtex4}
\usepackage{mathtools}
\usepackage{bm}
\usepackage{dsfont,amsthm,amsbsy}
\usepackage{verbatim}
\usepackage{amssymb}
\usepackage{amsmath}
\usepackage{bbm}
\usepackage{graphicx}
\usepackage{epstopdf}
\usepackage{subfigure}
\usepackage{natbib}
\usepackage{epsfig}
\usepackage{amsfonts}
\usepackage{mathrsfs}
\usepackage{sidecap}
\usepackage{lipsum}
\usepackage[toc,page,title,titletoc,header]{appendix}
\usepackage[colorlinks,linkcolor=blue,citecolor=blue,anchorcolor=blue,urlcolor=blue]{hyperref}
\usepackage{hyperref}
\usepackage{resizegather}
\usepackage{tikz}
\usepackage{float}
\usepackage{mathbbol}

\usepackage[normalem]{ulem}
\usepackage{cancel}
\usepackage{upgreek}

\usepackage{dblfloatfix}
\usepackage{graphicx}  
\usepackage{dcolumn}   
\usepackage{bm}        
\usepackage{amssymb}   
\usepackage{amsmath}
\usepackage{lipsum}
\usepackage{xcolor}
\usepackage{ulem}
\usepackage{braket}

\begin{document}

\title{From time crystals to time quasicrystals:\\ Exploring quasiperiodic phases in transverse field Ising chains}
\author{Davood Marripour}
\affiliation{Department of Physics, Institute for Advanced Studies in Basics Sciences (IASBS), Zanjan 45137- 66731, Iran}
\author{Jahanfar Abouie}
\email{jahan@iasbs.ac.ir}
\affiliation{Department of Physics, Institute for Advanced Studies in Basics Sciences (IASBS), Zanjan 45137- 66731, Iran}

\date{\today}

\begin{abstract}
Time quasicrystals (TQCs) represent a compelling extension of the concept of time crystals (TCs). While TCs break discrete time-translation symmetry by exhibiting a periodic response at a subharmonic of the driving frequency, TQCs display a more complex temporal order. They respond at multiple incommensurate frequencies, values that are not integer multiples of the fundamental driving frequency, resulting in quasiperiodic dynamics.
In this work, we investigate the emergence of a TQC in a disordered quantum Ising chain subjected to a quasiperiodic transverse field. Using exact diagonalization, we find that the transverse magnetization exhibits quasiperiodic oscillations which persist over extended prethermal timescales before eventual decay. This indicates that the TQC exists as a long-lived, prethermal dynamical phase rather than a true equilibrium state. We further assess the robustness of this prethermal TQC against interactions and driving imperfections, confirming its stability under realistic experimental conditions.
Finite-size analysis reveals that the prethermal TQC lifetime exhibits minimal dependence on system size.
Additionally, we explore the emergence of TQCs in the same chain under symmetric sampling of exchange couplings. Our results demonstrate that the TQC phase is highly sensitive to both the choice of coupling distribution and the values of the driving frequencies.
These findings highlight promising experimental prospects for realizing TQCs in cold atomic systems and quantum simulators, providing valuable insights into their stability, dynamical properties, and potential for exploring novel non-equilibrium quantum phases.

\end{abstract}

\maketitle
\section{Introduction}

Time crystals (TCs) represent a fascinating phase of matter that emerges from the spontaneous breaking of time translational symmetry (TTS) in periodically driven quantum systems \cite{r1,r2,r3,r4,r5,r6,Yousefjani1,Yousefjani2}. These systems are typically subjected to external periodic perturbations, allowing them to maintain coherence and exhibit long-lived states characterized by discrete time translational symmetry breaking. In this context, preventing heating is crucial to avoid a transition to infinite temperature, which would lead to the loss of any temporal order \cite{r7,r8}.

In disordered closed quantum systems, many-body localization (MBL) acts as a stabilizing mechanism against heating \cite{r99,r9}. MBL prevents the system from thermalizing by confining it to a localized region of the Hilbert space, thereby violating the eigenstate thermalization hypothesis (ETH). As a result, the quantum order is preserved over extended time scales, as the system retains its initial state within its eigenstates, shielded from thermalization processes \cite{r10,r11,r12,Pr12,rr12,rp12}.

Periodically driven Floquet systems typically exhibit non-conservation of energy due to continuous energy absorption from external fields. However, the presence of Floquet MBL alters this dynamic. By breaking the ETH, Floquet MBL prevents the system from reaching thermal equilibrium, leading to the emergence of nonequilibrium phases, including discrete TC phases \cite{r13,eqr13,r14,r15}.

In discrete TC phases, the dynamics of observables in the system are characterized by periodic responses that differ from the driving period. Specifically, for a Hamiltonian that oscillates with period $T$, the response functions of the system will exhibit oscillations with a period $nT$ (where $n \ge 2$) \cite{r8}.

Time quasicrystals (TQCs) represent a fascinating extension of the concept of time crystals (TCs). Unlike TCs, which break discrete time-translation symmetry by exhibiting a periodic response at a subharmonic frequency (a multiple of the driving frequency), TQCs display a more complex form of temporal order. They respond at multiple incommensurate frequencies, frequencies that are not integer multiples of the fundamental driving frequency, leading to quasiperiodic behavior in time.

TQCs arise from quasiperiodic driving of a system where the time-dependent Hamiltonian $H(t)$ incorporates irrationally related frequencies. This leads to quasiperiodic dynamics, resulting in a non-repetitive evolution of the system's state that densely explores a multi-dimensional toroidal phase space, without revisiting previous configurations. Such quasiperiodicity facilitates intricate dynamical behavior, including complex energy level spacings and dense trajectories in phase space \cite{rr17,dumitrescu2018,else2020}.

Recently, much attention has been devoted to the systems under quasiperiodic driving protocols \cite{verdeny2016,nandy2017,dumitrescu2018,giergiel2019,ray2019,zhao2021,pre13,else2020,mukherjee2020,FTQ,FTS,rr15,r17,r18,pre18,pre19}. In a recent proposal, it has been demonstrated that discrete TQCs can be realized in a Rydberg atomic chain by coupling two discrete TCs, where associated two external driving frequencies have the maximum incommensurability. In this system, the quasiperiodic response emerges due to interaction-induced synchronization between the subsystems \cite{pre24}.

It has been established that quasiperiodic driving ultimately leads to thermalization toward a featureless infinite-temperature state, but this process occurs after a long timescale known as the prethermal time \cite{dumitrescu2018}. This behavior is analogous to the prethermalization observed in high-frequency periodic driving, where the system first relaxes into a long-lived metastable state before eventual thermalization \cite{pre1,pre2,pre3,pre4,pre5,pre6,pre7,pre8,pre9,pre10}.

While rigorous results regarding heating rates are available for continuously varying quasiperiodic driving \cite{else2020}, there are also findings pertaining to discrete quasiperiodic driving protocols, such as those based on Fibonacci \cite{dumitrescu2018} and Thue-Morse sequences \cite{pre13,zhao2021}. Nonetheless, the general behavior under arbitrary quasiperiodic driving remains largely unexplored and requires further investigation.

The time evolution of entanglement entropy is a crucial parameter for distinguishing different dynamical regimes in driven many-body systems. In the MBL phase, the entanglement entropy grows logarithmically with time, $S(t) \sim \log t$, reflecting the slow dephasing dynamics and the absence of thermalization. In contrast, thermalizing systems exhibit a linear growth, $S(t) \sim t$, indicative of rapid entanglement spreading and eventual thermal equilibrium\cite{dumitrescu2018,pre20}. Under quasiperiodic driving, the entropy exhibits sublinear, power-law growth, $S(t) \sim t^{\alpha}$ with $\alpha<1$, characteristic of the prethermal regime and suggestive of a long-lived, non-thermal quasi-stationary state \cite{pre21,pre22}. 

In this study, we investigate the disordered ITF model. Recent research has demonstrated that the clean ITF model can exhibit a discrete TC phase driven by global periodic kicks generated by a transverse magnetic field, without the need for Hamiltonian quenching or long-range interactions \cite{Yu2019}. Building on this, we analyze the dynamical properties of the disordered ITF chain under periodic magnetic driving. Our results reveal the presence of a robust discrete TC phase in the disordered system. We further assess the stability of this phase against external perturbations and imperfections in spin rotations, finding it to be remarkably resilient. Using exact diagonalization, we observe that increasing the chain length enhances the stability of the discrete TC.

Additionally, we explore a prethermal TQC phase within a quasiperiodically driven ITF model. By leveraging prethermalization mechanisms, we demonstrate the emergence of a long-lived phase characterized by magnetization oscillating at multiple incommensurate frequencies. This TQC phase exhibits strong coherence despite interactions and imperfections in the quasiperiodic driving, persisting over extended timescales before eventual thermalization. Furthermore, we examine the manifestation of TQCs in disordered ITF chains with exchange couplings randomly drawn from a symmetric distribution. In this context, the system can reside in a spin glass phase. By examining various distributions of the exchange couplings $J_i$ and different driving frequencies, we analyze the conditions under which the TQC appears in the spin glass phase.

This paper is structured as follows: In Section \ref{Sec:TC}, we introduce the ITF spin chain, which is periodically driven by an external magnetic field, and analyze its magnetization in various directions, demonstrating the existence of a discrete TC phase. We further investigate the robustness of this discrete TC phase against perturbations in Section \ref{Sec:TC-Stability}. Section \ref{Sec:TQC} provides an overview of the definition and general properties of a TQC phase. In Section \ref{Sec:TQC-ITF}, we incorporate a sinusoidal driving field into the disordered ITF chain, revealing the existence of a TQC phase and exploring its stability under mechanism of prethermalization. Section \ref{Sec:TQC-SG} discusses the emergence of a TQC phase within an ITF chain exhibiting a spin glass phase. Finally, we summarize our findings in Section \ref{Sec:Conclud}.

\section{TC in disordered ITF model}\label{Sec:TC}

Let's consider a spin-$1/2$ Ising chain subjected to a transverse magnetic field (ITF) and periodically driven by an external magnetic field. The spin-spin interactions in this system are described by the following Hamiltonian:
\begin{equation}
	H(t)=
	\begin{cases}
		\hat{H}_{1}=g\sum\limits_{i}\hat{\sigma}_{i}^{x},\quad   0\le t_1<t^\prime\\
		\hat{H}_{2}=\sum\limits_{i}(-J_{i}\hat{\sigma}_{i}^{z}\hat{\sigma}_{i+1}^{z}+h_{i}\hat{\sigma}_{i}^{x}),\quad t^\prime<t_2\le T
	\end{cases}
	\label{Eq:itf-Hamiltonian} 
\end{equation}
Here, $t_1$ and $t_2$ are two dimensionless time scales satisfying $T=t_1+t_2$, where $T$ is a period indicating the time translational symmetry of the Hamiltonian, i.e. $\hat{H}(t)=\hat{H}(t+T)$.  In Eq. (\ref{Eq:itf-Hamiltonian}), $g$ is a static transverse magnetic field (set to $g=\frac{\pi}{2}$), $J_i$ is site-dependent exchange coupling, randomly chosen from the interval of $[\frac{1}{2}J,\frac{3}{2}J]$ with $J>0$, and $h_i$ is a site-dependent magnetic field, randomly chosen from the interval of $[-h,h]$, where $h$ is an arbitrary positive value. The Hamiltonian $\hat{H}_{2}$ possesses the global $Z_{2}$ symmetry with the Ising parity operator $\hat{P}=\prod_{i=1}^L \hat{\sigma}_{i}^{x}$, which rotates all spins around the $x$ axis\cite{A berif}.

In the clean system ($\delta J=J_i-J=0$ and $\delta h=h_i-h=0$), $\hat{H}_2$ reduces to the uniform ITF Hamiltonian. In this case, for $h\gg J$, the system is in the paramagnetic (PM) phase, where the ground state is gapped and nondegenerate, and all spins are aligned with the
transverse field $h$, illustrated as $|\rightarrow\rightarrow\rightarrow\dots\rangle$. In this phase the ground state has no long-range order, the magnetization $m^z$ is vanishing, and spin-spin correlations decay exponentially by increasing spin-spin separation distance. 
As the exchange coupling strength $J$ is increased, it promotes the alignment of spins along the $z$ direction, leading to the breaking of Ising symmetry at the critical point $h=J$, where a phase transition occurs from the PM phase to a FM phase. In the FM phase, the ground state is characterized by either $|\uparrow\uparrow\dots\rangle$ or $|\downarrow\downarrow\dots\rangle$, resulting in a finite magnetization and a definitive breaking of the Ising symmetry.

The situation is different for finite size chains. The system can tunnel between the two degenerate eigenstates at finite size, and consequently the two lowest lying eigenstates of the system are the following $Z_2$ symmetric states: 
$|\pm\rangle=\frac{1}{\sqrt{2}}[|\uparrow\uparrow\dots\uparrow\rangle\pm|\downarrow\downarrow\dots\downarrow\rangle]$. These states are the eigenstates of the Ising parity operator $\hat{P}$ with the eigenvalues $\pm 1$, they are long-range correlated with vanishing magnetization \cite{r7}.

At finite temperatures, the Mermin-Wagner theorem asserts that one-dimensional systems with discrete symmetries cannot undergo spontaneous symmetry breaking, as thermal fluctuations in these systems are strong enough to disrupt long-range order, resulting in a finite density of excitations\cite{mermin-wagner}. 
In the FM phase, these excitations manifest as domain walls. The energy associated with these domain walls is typically independent of their position along the chain, allowing them to move freely. This mobility of domain walls contributes to the destabilization of any potential long-range order, thus reinforcing the conclusion that spontaneous symmetry breaking cannot be realized in one-dimensional systems at finite temperatures.

The excited states can be described as superpositions of states featuring finite densities of domain walls distributed across various spatial locations. These delocalized and fluctuating domain walls introduce perturbations that disrupt long-range spin correlations at any finite temperature.

The introduction of disorder, characterized by a non-clean regime with $h_i \in [-h,h]$ and $J_i \in [J/2, 3J/2]$, leads to significant modifications in the system's physical properties. 
In the absence of external magnetic fields, the system resides in a FM phase characterized by a non-zero magnetization $m^z$, as all exchange couplings are positive. Upon the application of a weak magnetic field, specifically when $h\ll J$, domain walls emerge along the chain. For a single domain wall positioned at $d$, the excited state of the system can be expressed as $|d\rangle=|\uparrow\uparrow\dots \uparrow\downarrow_{_d}\uparrow\uparrow\dots\rangle$.
In contrast to the clean system, where a thermally activated domain wall can move freely, in the presence of disorder, the domain wall becomes immobilized due to the energy cost associated with its dispersion throughout the chain. 
As the magnetic field increases, the system can sustain a finite density of domain walls, all of which remain pinned by the disorder. It has been demonstrated that for $h<J$, the system exhibits characteristics of the many-body localized (MBL) phase within the FM phase, which remains stable in the presence of perturbations such as interactions\cite{r23}.

We now examine the potential for spontaneous breaking of time translational symmetry in the Floquet ITF model. The time evolution of the system, as described by equation (\ref{Eq:itf-Hamiltonian}), is governed by the Floquet operator given by:
\begin{equation}
	\hat{U}_{\rm F}=\exp(-it_2\hat{H}_{2})\exp(-it_{1}\hat{H}_{1}).
	\label{Eq:Time-Evol-Operat}
\end{equation}
With $t_1= t_2=1$, the total driving period of the Hamiltonian is $T = 2$, and accordingly, the drive frequency is $\omega = \frac{2\pi}{T} = \pi$. Also the Floquet operator is simplified to:  
\begin{equation}
	\hat{U}_{\rm F}=\exp(-i\hat{H}_{2})\hat{X},~~~
	\hat{X}=\prod\limits_{i}(-1)^i\hat{\sigma}_{i}^{x}.
\end{equation}
Here, the first term captures the time evolution over the interval $1< t_2\le T$ under the Hamiltonian $\hat{H}_2$, while the second term represents a spin rotation about the $x$-axis. Through this formalism, we aim to determine the conditions under which time translational symmetry may undergo spontaneous breaking in the context of the FITF model.

Consider the system initially in a fully polarized state along the $z$ axis, represented as $|\psi_0\rangle=\otimes_{i=1}^L |\uparrow_i\rangle$, where $L$ is the number of sites. The stroboscopic dynamics of the system are obtained by iteratively applying the Floquet time evolution operator, yielding the system's state at discrete times $t=nT$, where $n$ denotes the number of driving periods, along with the time dependence of various order parameters.
We analyze the temporal evolution of the order parameters, namely the magnetizations along the $x$, $y$, and $z$ axes, defined as 
\begin{equation}
m^{x,y,z}(t)=\left\langle\frac{1}{L}\sum_i^L\langle\psi(t)|\hat{\sigma}_i^{x,y,z}|\psi(t)\rangle\right\rangle, 
\label{eq: mag-def}
\end{equation}
where $|\psi(t)\rangle$ represents the system's state at time $t$, and $\langle \dots \rangle$ denotes the ensemble average over different random realizations.

\begin{figure}
	\includegraphics[scale=0.55]{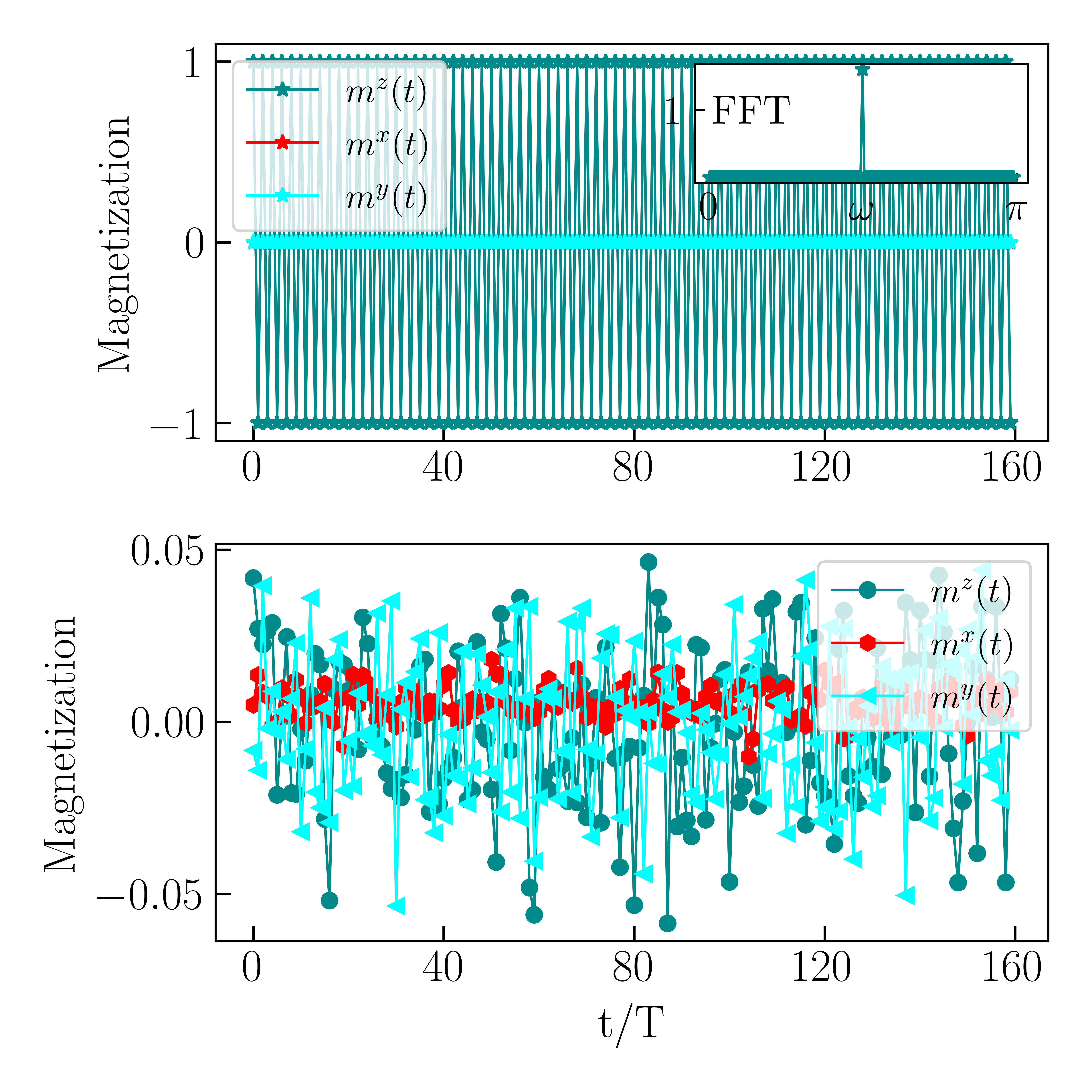}
	\caption{Time evolution of the magnetizations along $x, y$, and $z$ directions for a chain of 12 sites.
For $J=5.5$ and $h=0.3$ (up), the magnetization along the $z$ axis exhibits periodic oscillations with a period of $2T$, while the $x$ and $y$ components remain zero. The inset plot illustrates the fast Fourier transform (FFT) of $m^z(t)$, highlighting a pronounced peak at $\omega=\pi/2$, which signifies the periodic nature of the magnetization. Conversely, for $J=0.3$ and $h=4.5$ (down), the magnetization oscillations are characterized by a lack of periodicity and stability. The peak value in the FFT is multiplied by 0.01, and this adjustment is applied to all other plots as well.}
	\label{Fig:mag-TC}
\end{figure}

In the presence of disorder, it is essential to compute ensemble averages of the magnetizations across various random realizations. By averaging over 150 random realizations, we derive the magnetizations for 160 Floquet cycles. Our findings indicate that when $\bar{J}\gg\bar{h}$ ($\bar{h}$ and $\bar{J}$ represent the mean values of the random fields $h_i$ and the random couplings $J_i$, respectively), the magnetization $m^z(t)$ exhibits oscillatory behavior around zero with a period of $2T$ (refer to Fig. \ref{Fig:mag-TC}). The periodic oscillations of $m^z(t)$, with a period that is double that of the Hamiltonian, suggest a breakdown of time translational symmetry. This phenomenon is indicative of discrete TC, provided that the observed oscillations remain stable in the thermodynamic limit ($L\rightarrow\infty$) and are resilient against perturbations such as interactions.

We also derive the time dependence of all components of the magnetization in the PM phase, where $\bar{J}<\bar{h}$. Our findings, presented in Fig. \ref{Fig:mag-TC}-(down), indicate that while the magnetizations are nonzero, their oscillations exhibit a lack of periodicity and stability.

\section{Stability of the discrete TC phase}\label{Sec:TC-Stability}

In this section, we investigate the stability of the discrete TC in the presence of various external perturbations. We will demonstrate that the introduction of diverse perturbations, including interactions and imperfect rotations (deviations from perfect rotation), does not compromise the TC order.

\subsection{Stability of the discrete TC phase under interactions}

In this part, we examine the stability of the discrete TC phase under the influence of exchange interactions along the $x$-axis. In this context, the second term of the Hamiltonian in Eq. (\ref{Eq:itf-Hamiltonian}) is modified as follows:
\begin{equation}
\hat{H}_{2}=\sum\limits_{i}(-J_{i}\hat{\sigma}_{i}^{z}\hat{\sigma}_{i+1}^{z}+h_{i}\hat{\sigma}_{i}^{x})+\lambda\sum_{i}\hat{\sigma}_{i}^{x}\hat{\sigma}_{i+1}^{x},
\label{Eq:H2-int}
\end{equation}
where $\lambda$ denotes the strength of the interaction between nearest neighboring spins located at sites $i$ and $i+1$.

To elucidate the origin of the interaction represented by the $\lambda$-term, we can map the spin Hamiltonian (\ref{Eq:H2-int}) onto a fermionic Hamiltonian. By applying the transformation $\sigma_i^x \leftrightarrow \sigma_i^z$ and utilizing the Jordan-Wigner (JW) spin-fermion transformations\cite{MB18}, we observe that the $\lambda$-term transforms into $\lambda c_i^\dagger c_i c_{i+1}^\dagger c_{i+1}$, where $c_i$ ($c_i^\dagger$) are the annihilation (creation) operators for fermions at site $i$. This expression clearly indicates the interaction between two fermions located at sites $i$ and $i+1$.

\begin{figure}
	\includegraphics[scale=0.35]{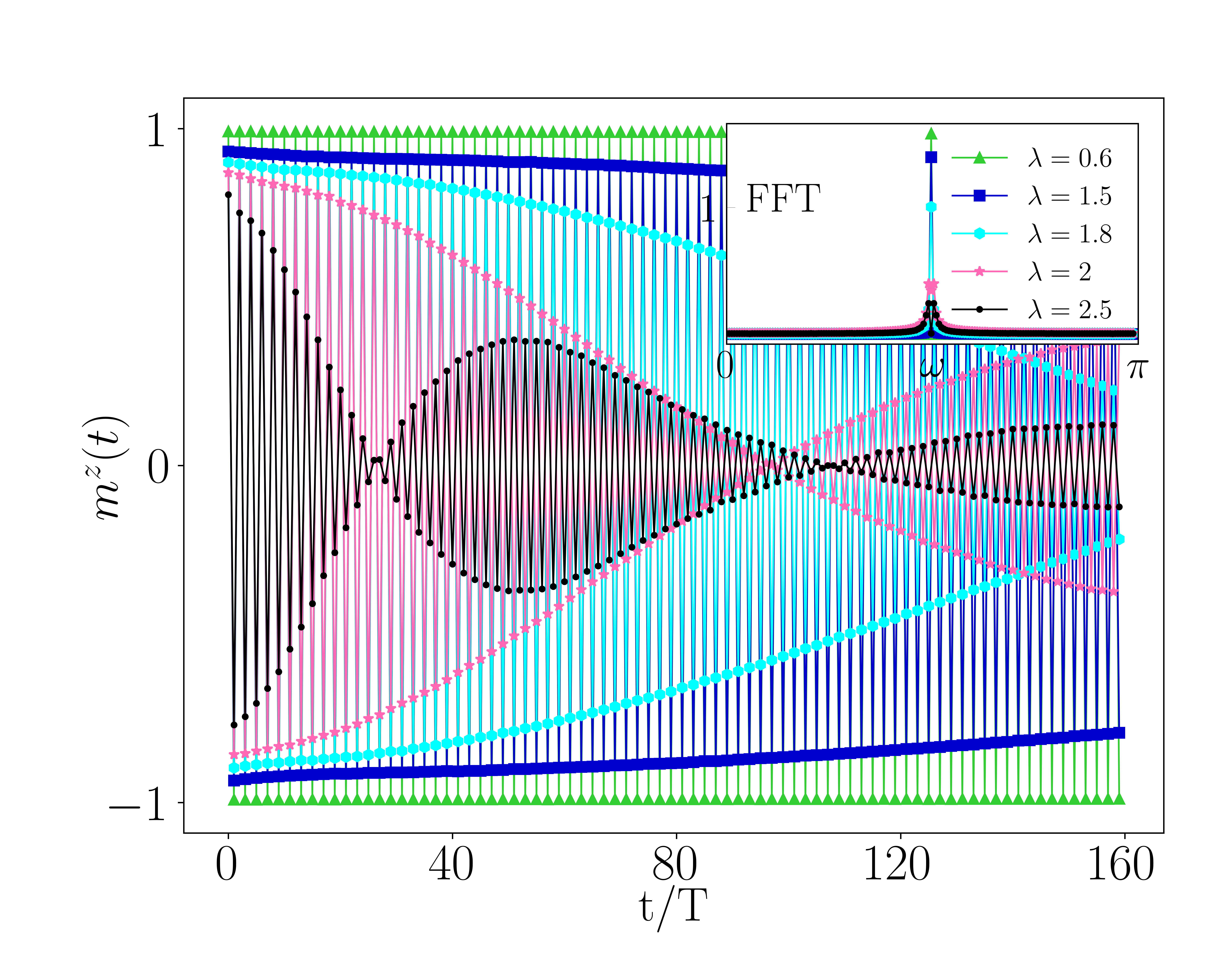}
	\caption{Time evolution of the magnetization along the $z$-direction for a chain of length $L=12$ subjected to an interaction $\lambda$ with varying strengths. The exchange couplings are randomly selected from the interval $[J/2, 3J/2]$ with $J = 5.5$, while the magnetic fields are drawn from the range $[-h, h]$ with $h = 0.3$. 
The envelope of the magnetization depends on the strength of the interaction $\lambda$. Increasing $\lambda$, leads to the instability of the oscillations with time period of $2T$, and consequently the destruction of the TC phase. FFT of the magnetization $m^z(t)$ exhibits a subharmonic peak at $\omega=\pi/2$. 
Increasing $\lambda$ results in a splitting of the Fourier peak, reflecting the beating of magnetization oscillations.}
	\label{Fig:mz-dif-lambda}
\end{figure}
In Fig. \ref{Fig:mz-dif-lambda}, we present the time evolution of the magnetization $m^z$ in the ferromagnetic many-body localized (FM-MBL) phase for varying interaction strengths. The observed time dependence of $m^z(t)$ reveals significant oscillatory behavior, characterized by a period of $2T$, superimposed on an oscillatory-decaying envelope that is dependent on the interaction strength $\lambda$. Specifically, for a finite-size system, the envelope exhibits a prolonged time period at low interaction strengths, and diminishes as $\lambda$ increases. 
Furthermore, we have computed the Fourier transform of $m^z(t)$ as a function of frequency for various $\lambda$ values.
For small values of $\lambda$, within the observed time window $t/T\in[0, 160]$, the magnetization oscillations remain smooth and exhibit no significant decay over time. The FFT analysis reveals a clear subharmonic peak at $\omega=\pi/2$. As $\lambda$ increases, this peak begins to split due to interference effects in the magnetization oscillations, eventually disappearing at a critical value $\lambda_c$, signaling the complete destruction of the TC phase. However, by extending the observation window, the decay of magnetization oscillations becomes more apparent over longer timescales, resulting in a double peak in the FFT even for small $\lambda$ values.

\begin{figure}
	\includegraphics[scale=0.35]{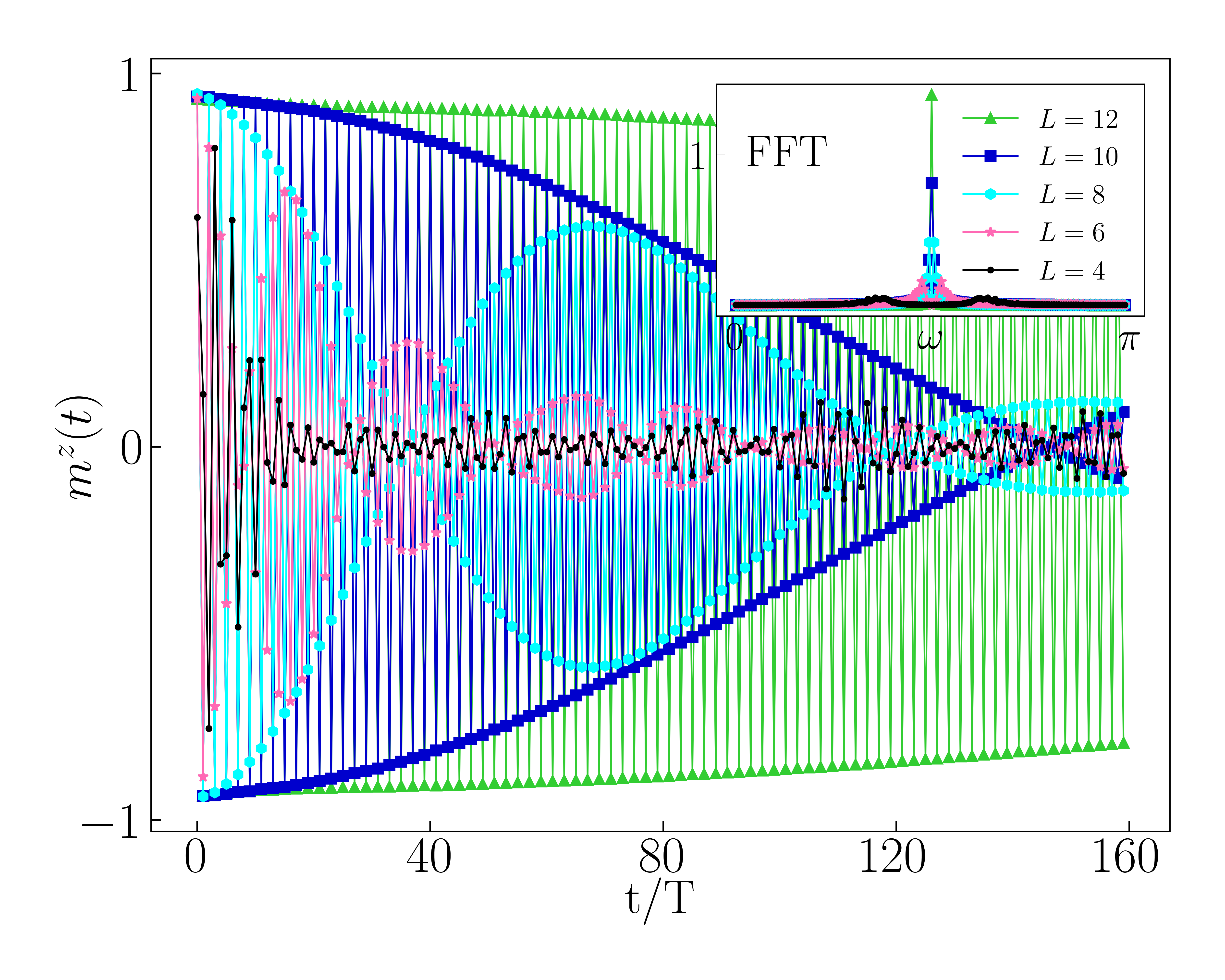}
	\caption{Time evolution of the magnetization along the $z$-direction for chains of varying lengths subjected to an interaction strength of $\lambda=1.5$. The exchange couplings are randomly selected from the interval $[J/2, 3J/2]$ with $J = 5.5$, while the magnetic fields are drawn from the range $[-h, h]$ with $h = 0.3$. As the system size increases, a significant enhancement in the stability of the magnetization oscillations is observed, characterized by a time period of $2T$. In the inset, the FFT of the magnetization $m^z(t)$ is presented. For smaller systems, such as with length $L=4$, the FFT, illustrated by the black symbols, reveals no sharp peak, indicating the absence of stable oscillations in $m^z(t)$. Conversely, as the system size increases, a pronounced peak emerges at the frequency $\omega=\pi/2$, signifying the establishment of stable oscillations in the magnetization.}
	\label{Fig:mz-dif-L}
\end{figure}
The observed oscillatory-decaying envelope in the magnetization can be attributed to finite-size effects. To substantiate this assertion, we present in Fig. \ref{Fig:mz-dif-L} the time evolution of the magnetization $m^z(t)$ at a coupling strength $\lambda<J$ for chains of varying lengths. Notably, as the chain length increases, the period of the envelope oscillations also increases, suggesting that in the limit of infinite chain length ($L\to\infty$), the envelope disappears, leaving only stable oscillations with a period of $2T$. Furthermore, in the inset of Fig. \ref{Fig:mz-dif-L}, we present the FFT of $m^z(t)$ for different system lengths. For smaller systems, the FFT reveals a double peak structure, reflecting the beating of magnetization oscillations. As the system size increases, these double peaks converge into a single peak at frequency $\omega=\pi/2$. Continued increases in system size result in a sharper peak that becomes increasingly concentrated around $\omega=\pi/2$.

\begin{figure}
	\includegraphics[scale=0.32]{absolute_lambda_tc}
	\centering
	\caption{(Top: The absolute value of the magnetization $m^z(t)$ for a chain of length $L=12$ for different values of $\lambda$, with a coupling constant $J=5.5$ and an external field $h=0.3$. The TC remains stable for interaction strengths weaker than a critical value of approximately $0.4 <\lambda_c < 0.5$. Bottom: Finite-size scaling of the critical interaction strength $\lambda_c$ as a function of $1/L$, where $L$ denotes the system size. The data points represent numerical values, and the solid line corresponds to a linear fit. This extrapolation suggests that $\lambda_c$ approaches a finite, nonzero value as $L\to\infty$. Specifically, the thermodynamic limit value of the critical interaction strength is estimated to be $\lambda_c(\infty) \approx 0.67 \pm 0.03$, indicating that the TC phase remains stable for interaction strengths below this threshold.}
	\label{Fig:stability_interaction}
\end{figure}
We can evaluate the stability of the discrete TC phase against interactions by examining how the absolute value of the magnetization, $|m^z(t)|$, varies over time. As shown in Fig. \ref{Fig:stability_interaction}-top, $|m^z(t)|$ remains constant over time for interaction strengths below a critical value $\lambda_c$, indicating that the TC phase is stable in this regime. The value of $\lambda_c$ depends on the system size; for a chain of length $L=12$, it lies approximately between 0.4 and 0.5.

To further understand the thermodynamic behavior, we performed a finite-size scaling analysis by plotting $\lambda_c$ as a function of $1/L$. As illustrated in Fig. \ref{Fig:stability_interaction}-bottom, $\lambda_c$ increases with system size, and a linear fit suggests convergence to a finite, nonzero value in the thermodynamic limit. This strongly indicates that the TC phase persists over a finite range of interaction strengths and is not merely a finite-size artifact, confirming its robustness in the thermodynamic limit. 

Refining the precise value of $\lambda_c$ necessitates analysis at larger system sizes. However, simulating chains of larger sizes presents substantial computational challenges, especially when a driving term complicates the dynamics. 
In this context, tensor network techniques, such as the time-evolving block decimation (TEBD) algorithm, offer a powerful solution. TEBD enables efficient representation and manipulation of quantum states, making it particularly well-suited for investigating quantum many-body systems. This approach is especially valuable when exact diagonalization becomes infeasible due to the exponential growth of the Hilbert space.

\subsection{Stability of the discrete TC under imperfect rotations}

We now examine the stability of the TC observed in the ITF model under conditions of imperfect rotations. Instead of executing a perfect spin-flip rotation (i.e., a rotation around the $x$-axis by an angle of $g=\frac{\pi}{2}$), we implement an imperfect rotation characterized by an angle of $g=\frac{\pi}{2}-\epsilon$, where $\epsilon$ is a small positive value ($\epsilon<1$). Consequently, the time evolution operator is expressed as in Eq. (\ref{Eq:Time-Evol-Operat}) with the time parameters set to $t_1=t_2=1$.

\begin{figure}
	\includegraphics[scale=0.35]{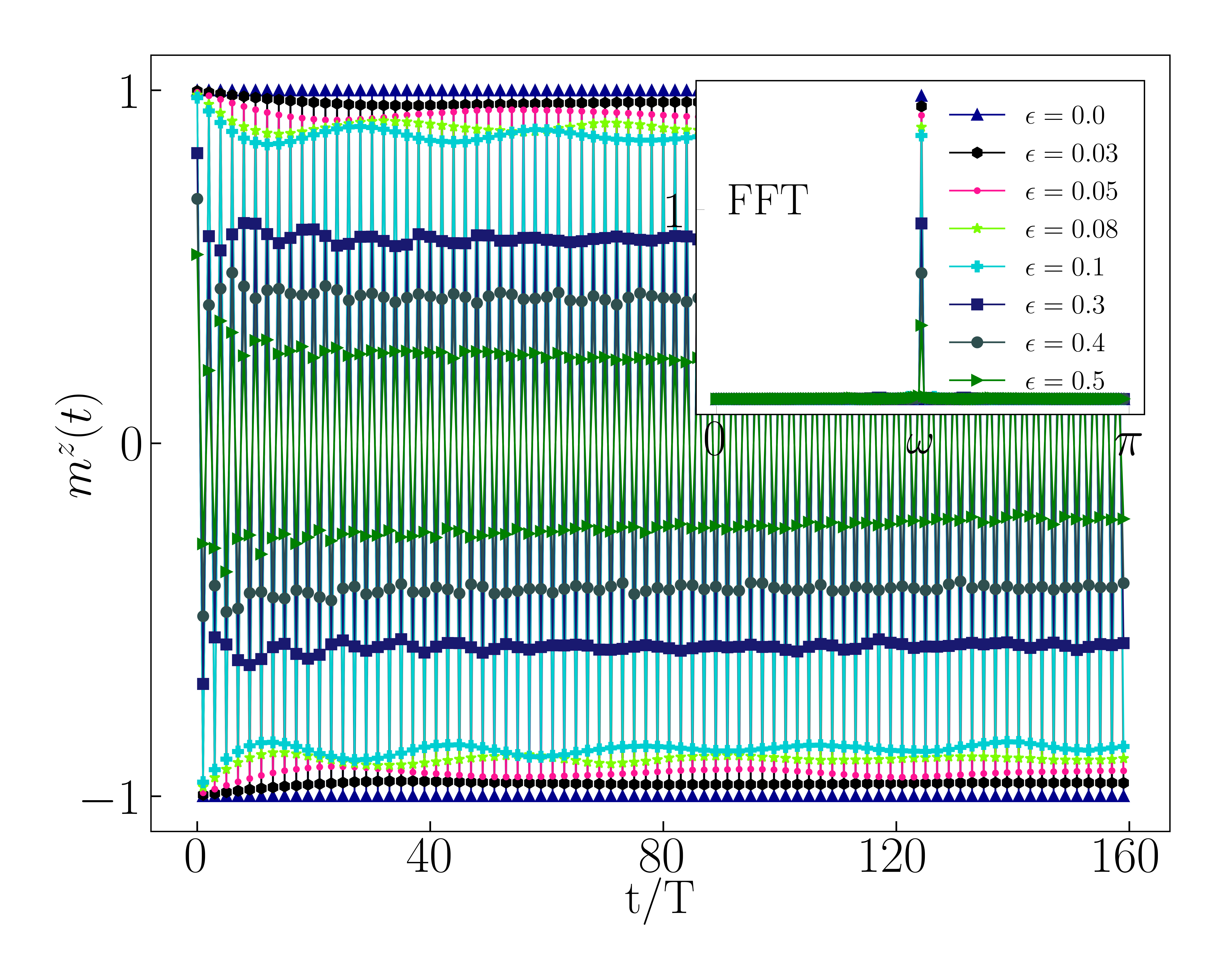}
	\centering
		\caption{Time evolution of the magnetization along the $z$-direction for various values of $\epsilon$ in a chain of length $L=12$ with coupling constant $J=5.5$ and external field $h=0.3$. The discrete TC remains stable against a finite degree of imperfect rotation. The inset shows the FFT of the magnetization. As imperfection increases, the peak at $\omega=\pi/2$ diminishes and ultimately vanishes for larger $\epsilon$, signaling the disappearance of the discrete TC phase.}
	\label{Fig:mz-dif-epsilon}
\end{figure}

The time-dependent behavior of the magnetization along the $z$-direction is depicted in Fig. \ref{Fig:mz-dif-epsilon}. In the presence of the imperfection parameter $\epsilon$, we observe that the magnetization continues to exhibit oscillatory behavior with a period of $2T$, which is twice the intrinsic period of the Hamiltonian. 
As $\epsilon$ increases, the amplitude of the magnetization oscillations decreases, as indicated by the declining peak value in the FFT of the magnetization $m^z(t)$ (see inset of Fig. \ref{Fig:mz-dif-epsilon}). Despite this reduction, the magnetization oscillations remain stable over time. However, further increases in $\epsilon$ lead to a continued decline in oscillation amplitude, ultimately destabilizing the discrete TC phase. For a chain length of $L=12$ (refer to Fig. \ref{Fig:mz-dif-epsilon}), the critical value for stable magnetization oscillations is approximately $\epsilon_c\approx 0.3$; beyond this point, the oscillations become unstable and their amplitude diminishes. 
Notably, as the chain length increases, the stability of the discrete TC phase improves. Establishing the exact value of critical $\epsilon$ will necessitate further computational simulations on larger chains, which falls outside the scope of this study.

\section{Time Quasicrystal (TQC): general definition}\label{Sec:TQC}

In this section, we delve into the disordered ITF chain influenced by a quasiperiodic driving field, showcasing its ability to exhibit a TQC phase. Before proceeding, we will first define the TQC phase and outline the conditions necessary for the emergence of a TQC within a model's phase diagram.

Let us consider a time-dependent Hamiltonian $\hat{H}(t)$ as follows\cite{verdeny2016}: 
 \begin{equation}
 	\hat{H}(t)=\sum_{\vec{n}}\hat{H}_{\vec{n}}\exp(i\vec{n}\cdot\vec{\omega}t),
 	\label{Eq:quasiperiod-Hamiltonian}
 \end{equation}
where $\hat{H}_{\vec{n}}$ represents a time-independent operator, $\vec{\omega}$ is a vector of dimension $d$ with components $\omega_1, \dots, \omega_d$, corresponding to the oscillation frequencies of different terms of the Hamiltonian, and $\vec{n}$ is a $d$-dimensional integer vector with components $n_1, \dots, n_d$. The summation is carried out over all integer values of $n_j$. 

When the frequencies $\omega_j$ in the Hamiltonian $H(t)$ are rationally related, the system exhibits periodic behavior, meaning that the dynamics of the system will repeat after a certain period, which can be expressed as the least common multiple of the periods associated with the individual frequencies. In contrast, when the frequencies $\omega_j$ are irrationally related, the Hamiltonian becomes quasiperiodic.
Quasiperiodicity implies that the system does not repeat in a simple manner; instead, its state evolves over time in a way that covers a multi-dimensional toroidal space without returning to a previous configuration.

In the context of periodically and quasiperiodically driven systems, the concepts of time translation symmetry breaking (TTSB) and quasiperiodic time translation symmetry breaking (QTTSB) provide a framework for analyzing the behavior of operators over time. 
For a periodically driven system with period $T$, TTSB occurs when the ensemble average of an operator $\hat{O}$ satisfies the condition $\langle \hat{O}(t+mT)\rangle=\langle \hat{O}(t)\rangle $ for integers $m\ge 2$. This indicates that the system exhibits a form of symmetry breaking where the observable properties of the system change over time in a different repeating manner.

On the other hand, in quasiperiodic systems, there is no single characteristic period due to the nature of their driving, which is typically defined by incommensurate frequencies. However, QTTSB can still be characterized through the frequency analysis of the system’s dynamics. Specifically, a quasiperiodic system demonstrates TQC behavior if there exists an operator $\hat{O}$ whose ensemble average satisfies the following Fourier transformation property \cite{verdeny2016}:
\begin{equation}
	\langle \hat{O}_t\rangle=\sum_{\vec{n}}\langle \hat{O}_{\vec{n}}\rangle\exp(i\vec{n}\cdot\vec{\tilde{\omega}}t),
	\label{Eq:quasiperiod-order}
\end{equation}
where $\vec{n}$ is a $d$-dimensional vector of integers, $\vec{\tilde{\omega}}$ is a $d$-dimensional vector with components $\tilde{\omega}_1, \dots, \tilde{\omega}_d$, where $\tilde{\omega}_j=\omega_j/p_j$. Here, $\omega_j$ are the driving frequencies, and $p_j$ are non-negative integers with at least one $p_i$ exceeding 1. This implies that the system exhibits a quasicrystalline ordering in time, characterized by multiple frequencies that do not lead to a simple periodic repetition but rather a complex structure reminiscent of quasicrystals in spatial dimensions\cite{FTS}.

We now investigate the emergence of a TQC phase within the disordered ITF chain. A promising approach involves replacing the periodically driven $\pi$-pulses with a quasiperiodic driving scheme. In this context, we examine the following time-dependent Hamiltonian:
 \begin{equation}
	\hat{H}(t)=\hat{H}_{ITF}(t)+\hat{H}_{drive}(t)+\frac{\Omega}{2}\sum_{i}\hat{\sigma}_{i}^{z}
	\label{Eq:ITF-Hamiltonian-quasi} 
\end{equation}
where,
 \begin{equation}
	\hat{H}_{\text{ITF}}(t)=h_1(t)\sum\limits_{i}[-J_{i}\hat{\sigma}_{i}^{z}\hat{\sigma}_{i+1}^{z}+h_{i}\hat{\sigma}_{i}^{x}],
	\label{Eq:ITF-Hamiltonian-MBL}
    \end{equation}
and
\begin{equation}
	\hat{H}_{drive}(t)=h_2(t)\sum\limits_{i}[\cos(\Omega t)\hat{\sigma}_{i}^{y}-\sin(\Omega t)\hat{\sigma}_{i}^{x}].
	\label{Eq:quasi-Hamiltonian-drive}
\end{equation} 
The time-dependent Hamiltonian in Eq. (\ref{Eq:ITF-Hamiltonian-quasi}) consists of three main terms: 

(1) ITF Hamiltonian $\hat{H}_{ITF}$ which describes the interactions in the system, including nearest-neighbor couplings and a transverse field. 

(2) Driving Hamiltonian $\hat{H}_{drive}$ which introduces a time-dependent driving force that oscillates sinusoidally. 
The driving terms involve both the $x$ and $y$-components of the spin operators, which can induce interesting dynamics in the system. 

(3) Additional magnetic field term $\frac{\Omega}{2}\sigma_i^z$, representing a static magnetic field in the $z$-direction.

The time-dependent coefficients $h_1(t)$ and $h_2(t)$ are switched on and off periodically, enabling a controlled temporal modulation of the system's evolution. Precisely defining $h_1(t)$ and $h_2(t)$ is essential for managing the system's dynamics, as their piecewise structure prevents overlap between the effects of the ITF and the driving terms. This separation allows for a clear, stepwise evolution, which can give rise to non-trivial dynamical behavior.

We specify $h_1(t)$ and $h_2(t)$ as follows:
\begin{equation}
	\begin{aligned}
		h_1(t) &= 
		\begin{cases}
			0, & \text{for } \frac{T}{4} < t < -\frac{T}{4} \\
			1, & \text{for } -\frac{T}{4} \le t \le \frac{T}{4}
		\end{cases}, \\
		h_2(t) &= 
		\begin{cases}
			\frac{\pi}{T}, & \text{for } \frac{T}{4} < t < -\frac{T}{4} \\
			0, & \text{for } -\frac{T}{4} \le t \le \frac{T}{4}
		\end{cases}.
	\end{aligned}
	\label{Eq:h1-h2}
\end{equation}
This construction ensures that, during the driving protocol, the system dynamically transitions through distinct Hamiltonian regimes in a well-defined sequence, thereby shaping its overall evolution.

According to Eq. (\ref{Eq:h1-h2}), the driving terms $h_2(t) \cos(\Omega t)$  and $h_2(t) \sin(\Omega t)$ in Eq. \eqref{Eq:quasi-Hamiltonian-drive}, are consistent with the definition of quasiperiodicity as specified in Eq. \ref{Eq:quasiperiod-Hamiltonian}. This consistency can be verified through their Fourier expansion, which is given by:
 \begin{equation}
h_2(t)\cos(\Omega t)=\sum_{n_1=-\infty}^{\infty}\sum_{n_2=\pm1}c_{n_1}e^{i(n_1\omega+n_2\Omega)t},
	\label{Eq:quasi-Hamiltonian-drive-expansion}
\end{equation}
where, $\omega=\frac{2\pi}{T}$ and $c_{n_1}$ are the coefficients of the expansion. 
The coefficients $c_{n_1}$ in the expansion will determine the specific nature of the quasiperiodicity, and the interplay between the various frequencies $n_1\omega+n_2\Omega$ can lead to complex dynamics\cite{FTS}.

\section{TQC in the disordered ITF model}\label{Sec:TQC-ITF}

The phenomenon of QTTSB can be characterized through the temporal evolution of the magnetizations $m^{x,y,z}=\langle\hat{\sigma}^{x,y,z}\rangle$. While quasienergy level statistics are typically employed to investigate the thermalization phase transition in periodic systems, as noted in \cite{r8}, our quasiperiodic system presents challenges in defining the Floquet operator and quasienergies.
In quasiperiodic systems, the absence of discrete TTS hinder the application of stroboscopic simplifications. Consequently, it is necessary to analyze the system's dynamics through the time evolution operator $U(t)=\mathcal{T}\exp(-i \int_0^t H(t') dt' )$, where $\mathcal{T}$ denotes time-ordering, and $H(t')$ is the quasiperiodic Hamiltonian, given by Eq. (\ref{Eq:ITF-Hamiltonian-quasi}).

For a chain of length $L=10$ with periodic boundary conditions, we employ exact diagonalization to compute the magnetization dynamics governed by the Hamiltonian presented in Eq. \eqref{Eq:ITF-Hamiltonian-quasi}.  The frequencies are fixed at $\omega=1$ ($T=2\pi$) and $\Omega=\sqrt{2}/4$, while the exchange couplings $J_i$ are randomly selected form the interval $[J/2, 3J/2]$, and the transverse fields $h_i$ are drawn from the interval $[0, h]$. The magnetization data are sampled at a time step of $\frac{T}{200}$. 

The evolution of the magnetizations are initiated from the initial state $|\psi_0\rangle=\otimes_i^L[|\uparrow_i\rangle+|\downarrow_i\rangle]$. 
By averaging over 200 random realizations, we obtain the time dependence of the magnetizations over a time window of $100 T$. 
The results, depicted in FIG. \ref{Fig:magnetization-quasi}, reveal that when $J\gg h$ the subharmonic response is evident in the Fourier decomposition of the drive as expressed in Eq. \eqref{Eq:quasi-Hamiltonian-drive-expansion}. This response distinctly indicates QTTSB, characterized by sharp peaks at $k\frac{\omega}{2}\pm\Omega$ for odd integer $k$. 
\begin{figure}
	\includegraphics[scale=0.19]{mag_tqc}
	\centering
	\caption{Top: Time evolution of the magnetization $m^x(t)$ for a chain of length $L=10$, with exchange coupling $J=5.5$ and transverse field $h=0.3$. The plot reveals periodic oscillations in $m^x(t)$, capturing the essential features of the quasiperiodic dynamics. The magnetization $m^x(t)$ exhibits an initial decay followed by the first plateau extending from $\tau_{\mathrm{pre}}$ to $\tau^{*}$. The persistence of coherent oscillations within this time interval highlights the stability of the prethermal regime, with its duration $t_{\mathrm{th}}=\tau^*-\tau_{\mathrm{pre}}$ defining the prethermal lifetime. Bottom: The FFT of $m^x(t)$ as a function of frequency $\omega$ exhibits sharp peaks at $k\omega/2 \pm \Omega$, where $k$ is an odd integer. These peaks are indicative of QTTSB and signal the emergence of a TQC in the system.}
	\label{Fig:magnetization-quasi}
\end{figure}

The magnetization amplitude exhibits a smooth decline punctuated by temporal plateaus (see the envelope in Fig.~\ref{Fig:magnetization-quasi}). This behavior, which arises from the transverse field in the Ising chain, contrasts with that of a chain subjected to a longitudinal field, where the magnetization remains relatively stable over comparable timescales \cite{FTS}. The observed decay of the envelope indicates that the stability of the TQC is limited; over sufficiently long times, the system heats up and approaches an infinite-temperature state \cite{dumitrescu2018, pre13, zhao2021}. This thermalization results from the quasiperiodic driving and occurs after a characteristic prethermal time, $t_{\mathrm{th}}$, which depends on the driving parameters \cite{dumitrescu2018}.

To quantify $t_{\mathrm{th}}$, we analyze the envelope of the magnetization $m^x(t)$. We identify two characteristic times from this envelope. The first, denoted $\tau_{\mathrm{pre}}$, marks the onset of the prethermal regime. It is defined as the time when the initial transient decay concludes and the magnetization enters its first nearly flat region (solid red line in Fig.~\ref{Fig:magnetization-quasi}). The second time, $\tau^{*}$, corresponds to the end of this first plateau, where the envelope begins its decay due to heating. We define the prethermal lifetime as $t_{\mathrm{th}} = \tau^{*} - \tau_{\mathrm{pre}}$, following the convention established in Ref.~\cite{Kemp2017,prethermal2021}. This definition facilitates systematic comparisons across different system sizes and driving conditions, aligning our analysis with the broader literature on prethermal phases.

The first plateau is identified as the period during which the normalized magnetization $m^x(t)$ maintains an amplitude within the interval $0.9 < m^x < 1$.

It is important to note that, in the Ising model subjected to a longitudinal magnetic field, a unitary transformation exists that maps the quasiperiodic Hamiltonian onto a periodic one. This transformation facilitates the transition from a TQC to a TC \cite{FTS}. 
However, such a transformation does not exist for the ITF model. Specifically, the transformation would need to convert the quasiperiodic drive into a periodic one while leaving the static Ising Hamiltonian unchanged. While this is possible for the Ising model in a longitudinal field, no similar transformation is available for the ITF chain because the transverse term in the Hamiltonian does not commute with the interaction term.

\subsection{Stability of the TQC phase in the ITF chain}

In this section, we will analyze the stability of the TQC phase when subjected to different perturbations. Prior to this, it is helpful to review how a non-equilibrium system thermalizes under various types of driving fields.

Local time-dependent Hamiltonians can only induce local rearrangements. In strongly disordered systems, such rearrangements require a nonzero energy cost and are typically nonresonant with the discrete harmonics of the periodic drive, thereby preserving MBL. In contrast, random driving, with its continuous frequency spectrum, can resonantly activate arbitrary local transitions, leading to energy absorption and eventual thermalization. Consequently, MBL is generally unstable under random driving. Quasiperiodic drives, which consist of multiple incommensurate discrete frequencies, create a noncontinuous but dense spectrum that can preserve MBL under typical conditions. However, over long times, they may lead to slow thermalization of the system. However, over long timescales, which diverge exponentially for weak or rapid drives, the system eventually thermalizes to infinite temperature. Nonetheless, slow relaxation processes can give rise to metastable dynamical phases, reminiscent of prethermal regimes \cite{dumitrescu2018,zhao2021}. 

The presence of a long-lived quasi-MBL regime with only slow decay suggests the emergence of transient phases unique to quasiperiodically driven systems. Analogous to prethermal state, this regime can be interpreted as a prethermal TQC phase. While the system does eventually thermalize, the process is sufficiently slow to allow the emergence of a TQC phase. The prethermalization lifetime depends on several parameters, notably the drive frequency, where higher frequencies prolong the prethermal regime. Importantly, under quasi-periodic driving, the heating dynamics are highly sensitive to the drive waveform smoothness; rectangular pulses markedly increase energy absorption and accelerate thermalization, whereas smoother waveforms such as sinusoidal drives substantially enhance the stability and duration of the prethermal state. Additionally, the drive amplitude critically influences the system’s energy uptake, with larger amplitudes expediting thermalization \cite{rr17,dumitrescu2018,zhao2021}.

Following, we systematically investigate the stability of the observed TQC phase within the disordered ITF chain, with a specific focus on the effects of interactions and imperfections.

\subsection{The stability of the TQC phase under interaction perturbations}

In order to investigate the stability of the TQC phase in the presence of interaction perturbations, following a similar approach as before, we incorporate the interacting $\lambda$ term into the ITF Hamiltonian from Eq. (\ref{Eq:ITF-Hamiltonian-quasi}) as:
 \begin{equation}
	\hat{H}_{\text{ITF}}(t)=h_1(t)\sum\limits_{i}[-J_{i}\hat{\sigma}_{i}^{z}\hat{\sigma}_{i+1}^{z}+h_{i}\hat{\sigma}_{i}^{x}+\lambda \hat{\sigma}_{i}^{x}\hat{\sigma}_{i+1}^{x}].
	\label{Eq:ITF-Hamiltonian-quasi-int}
\end{equation}
Given that the time-dependent coefficient $h_1(t)$ is periodically turned on and off within the time interval specified in Eq. (\ref{Eq:h1-h2}), the interaction perturbation is similarly introduced periodically into the system during this interval.

\begin{figure}
	\includegraphics[scale=0.19]{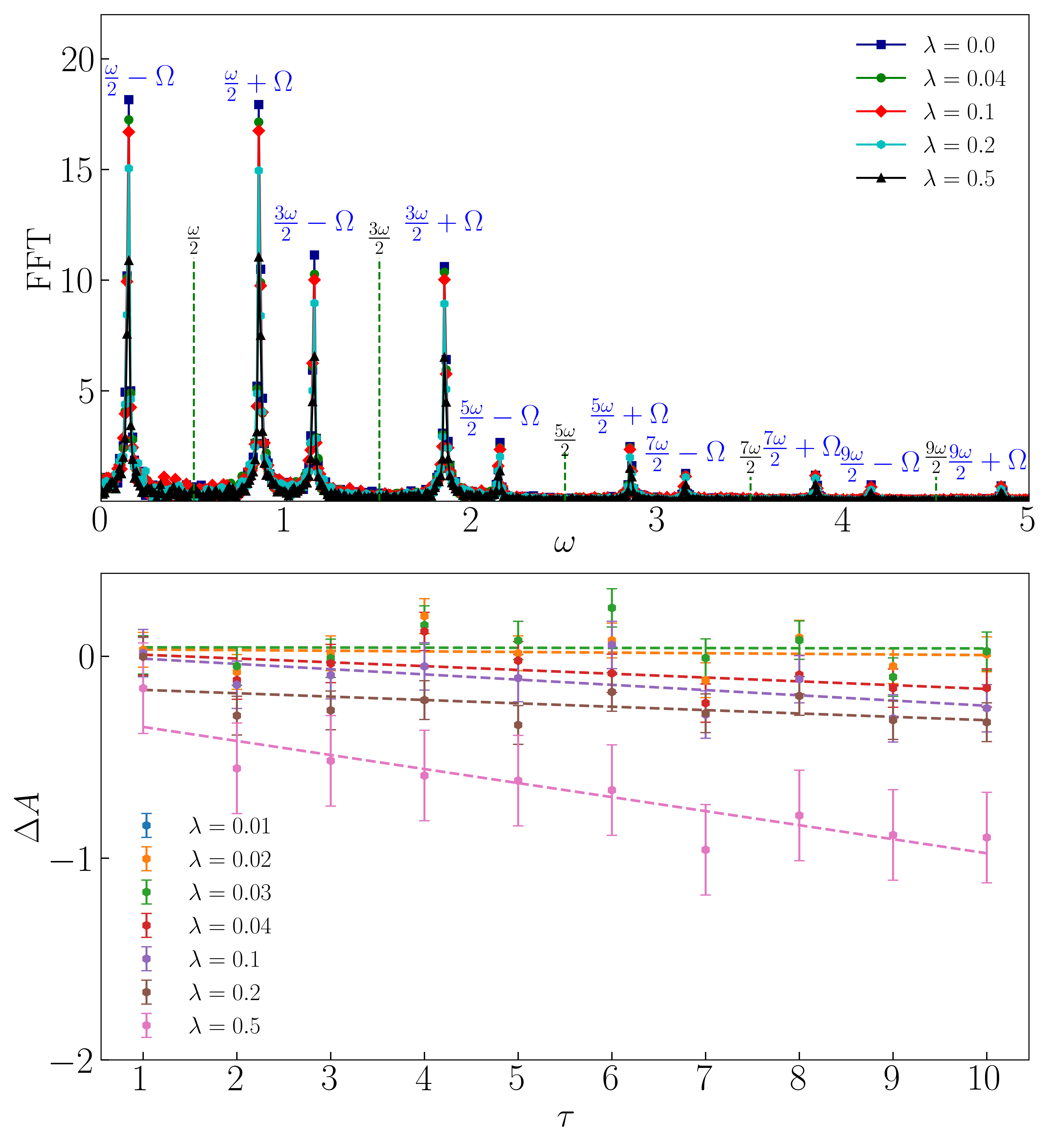}
	\centering
	\caption{ 
	Up: FFT analysis of the magnetization $m^x(t)$ in the TQC phase for various values of the interaction strength $\lambda$ (up). The data, derived from exact diagonalization of the Hamiltonian for a disordered ITF chain of length 8 with parameters $J=5.5$ and $h=0.3$, are averaged over 200 random realizations. FFTs are computed over a time window of $100T$. Observed peaks at $k\omega/2 \pm\Omega$  (where $k$ is an odd integer), demonstrate the resilience of TQC against interaction perturbations. Notably, as $\lambda$ increases, peak amplitudes decrease, ultimately vanishing at a critical value $\lambda_c$. Down: Variation of $\Delta A$ across different time windows ($\tau$) for various interaction perturbations. For a chain length of $L=8$, the TQC phase becomes unstable for interactions exceeding $\lambda_c=0.03$, as indicated by the observed decay at $\lambda>\lambda_c$. Error bars represent the standard deviation of the data, calculated across different time windows.}
	\label{Fig:FFT-quasi-int}
\end{figure}

The FFT analysis of the magnetization $m^x(t)$ is presented in Fig. \ref{Fig:FFT-quasi-int} for different values of the interaction strength $\lambda$. The prominent peaks observed at $k\omega/2 \pm\Omega$, where $k$ is an odd integer, suggest that the TQC phase demonstrates resilience to a range of interaction strengths.

Two key observations emerge from the FFT plot as a function of frequency across varying interaction strengths. Firstly, the positions of the peaks remain unaffected by the interaction, indicating that the fundamental frequencies associated with the magnetization oscillations are intrinsic to the system. This invariance implies that the overall periodic behavior of magnetization is robust against perturbations introduced by the interactions. It indicates that the system retains its underlying oscillatory nature, reflecting the translational symmetry of the system.

Secondly, the variation in the amplitude of the magnetization oscillations, illustrated by changes in the peak magnitudes, indicates that while the system's fundamental frequencies remain constant, the strength of the oscillations is influenced by interaction strength. This could suggest that stronger interactions suppress the magnetization dynamics, affecting how pronounced the oscillations are. 

The lack of smaller peaks at higher frequencies suggests that these frequencies are no longer representative of the system's dynamics. This can be interpreted in various ways: it may signify a damping effect due to heightened interactions, or it could indicate a transformation in the system's energy landscape, rendering higher frequency modes less significant. However, given that the amplitude of magnetization oscillations diminishes over time in the presence of interaction perturbations, the absence of smaller peaks at higher frequencies strongly points to a damping effect. 

As the parameter $\lambda$ increases, the peak amplitudes diminish and ultimately vanish at a critical value, $\lambda_c$. To determine $\lambda_c$, we employ the short-time Fourier transform on the magnetization $m^x(t)$. We segment the time interval of $100T$ into ten equal windows, performing a Fourier transform within each window to extract the amplitude of the first peak for varying values of $\lambda$. The difference in amplitude of the first peak at a non vanishing $\lambda$ and the amplitude at $\lambda=0$, denoted as $\Delta A$, provides valuable insight into the critical value $\lambda_c$. As shown in Fig. \ref{Fig:FFT-quasi-int}-(down), $\Delta A$ approaches zero for $\lambda$ values below $\lambda_c$, indicating that the TQC remains stable against interactions weaker than $\lambda_c$. For a chain of length $L=8$, the critical interaction value is approximately $0.03$, as illustrated in Fig. \ref{Fig:FFT-quasi-int}-(down).

\subsection{The stability of the TQC phase under imperfect rotations}

We now examine the stability of TQC under imperfect rotations. To this end, we modify the drive term of the Hamiltonian in Eq. (\ref{Eq:ITF-Hamiltonian-quasi}) as:
 \begin{equation}
	\hat{H}_{drive}(t)=h_2(t)\sum\limits_{i}[(1-\epsilon)\cos(\Omega t)\hat{\sigma}_{i}^{y}-\sin(\Omega t)\hat{\sigma}_{i}^{x}],
	\label{Eq:quasi-Hamiltonian_drive_imp}
\end{equation}
where the parameter $\epsilon$ quantifies the degree of imperfection in the spin rotations. The inclusion of the factor $(1-\epsilon)$ in front of the $\cos(\Omega t)$ term models the deviation from ideal rotation, effectively capturing the impact of imperfect control on the system's dynamics.

\begin{figure}
 	\includegraphics[scale=0.19]{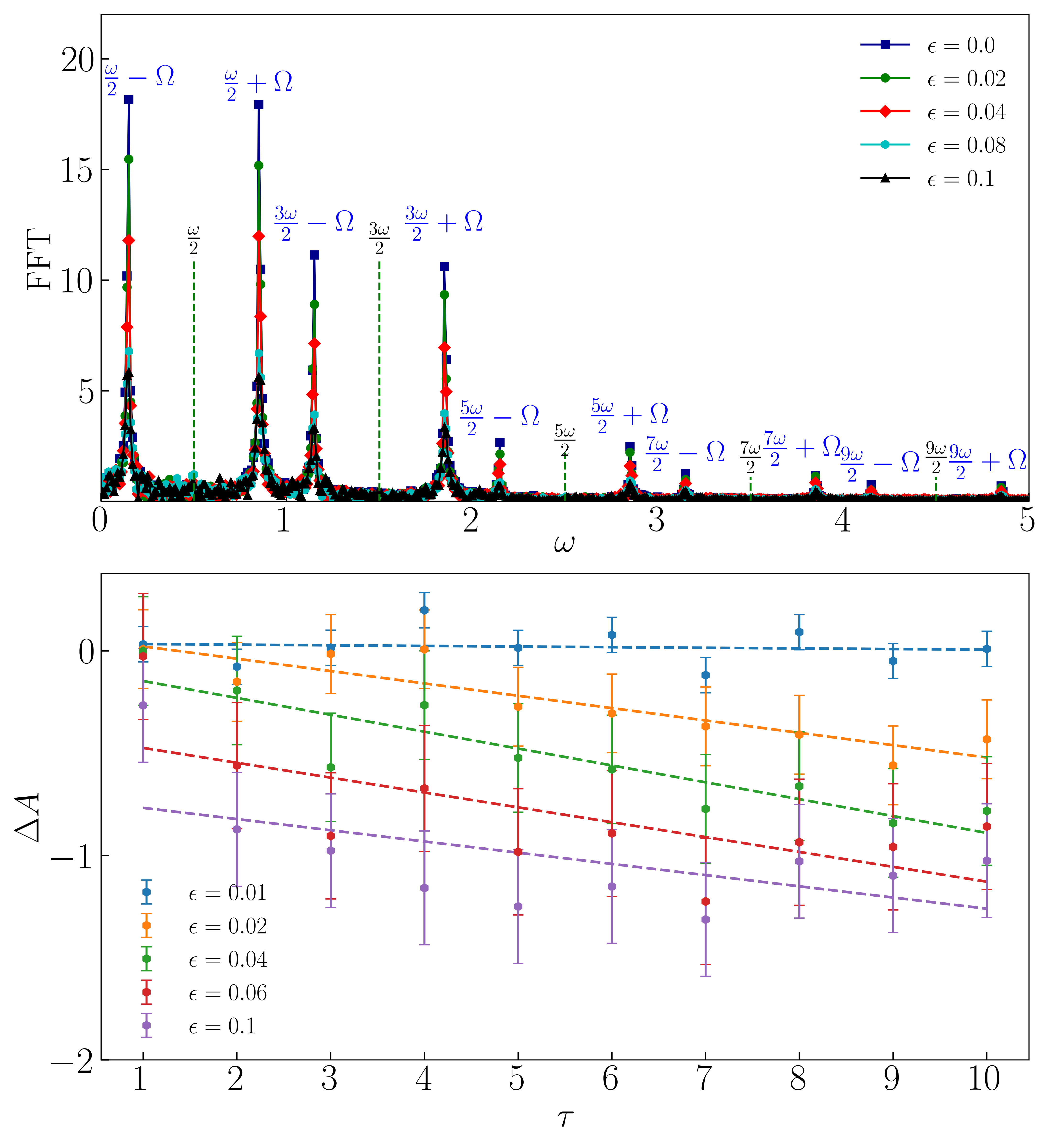}
 	\centering
 	\caption{Up: FFT analysis of the magnetization $m^x(t)$ in the TQC phase for various values of the imperfection $\epsilon$. The data, derived from exact diagonalization of the Hamiltonian for a disordered ITF chain of length 8 with parameters $J=5.5$ and $h=0.3$, are averaged over 200 random realizations. FFTs are computed over a time window of $100T$. Observed peaks at $k\omega/2 \pm\Omega$  (where $k$ is an odd integer), demonstrating the resilience of TQC against imperfections. Notably, as $\epsilon$ increases, peak amplitudes decrease, ultimately vanishing at a critical value $\epsilon_c$ ($\sim 0.01$ for a chain of length 8). Down: Variation of $\Delta A$ across different time windows ($\tau$) for various imperfections. For a chain length of 8, the TQC phase becomes unstable for imperfections exceeding $\epsilon_c=0.01$, as indicated by the observed decay at $\epsilon>\epsilon_c$.}
 	\label{Fig:FFT-quasi-imperfect}
 \end{figure}

The FFT of the magnetization $m^x(t)$ is illustrated in Fig. \ref{Fig:FFT-quasi-imperfect} for various values of the parameter $\epsilon$. The impact of imperfection is similar to the interaction perturbation. 
A key observation from the FFT of the magnetization $m^x(t)$ presented in Fig. \ref{Fig:FFT-quasi-imperfect} is that the presence of imperfections leads to a reduction in the amplitude of the subharmonic peaks without affecting their positions. Specifically, these subharmonic peaks appear at locations given by $k\frac{\omega}{2}\pm\Omega$ where $k$ is an odd integer, indicating that while the system's frequency structure remains stable, the strength of the response diminishes as $\epsilon$ increases. 

This stability of the peak locations suggests that our system can tolerate a certain degree of imperfections in the rotation, represented by $\epsilon$, without losing the essential characteristics that lead to a TQC phase. As such, even in the presence of imperfect rotations, the system retains its features and can continue to function robustly as a TQC.

To determine $\epsilon_c$, we study the behavior of the difference in the amplitude of the first peak at a non vanishing $\epsilon$ and the amplitude at $\epsilon=0$, denoted as $\Delta A=A(\epsilon\neq 0)-A(\epsilon=0)$, for a chain of length 8. As shown in Fig. \ref{Fig:FFT-quasi-imperfect}-(down), $\Delta A$ approaches zero for $\epsilon$s below $\epsilon_c=0.01$, indicating that the TQC remains stable against imperfections smaller than $0.01$. 

In summary, the findings indicate that the TQC phase can be maintained under the influence of rotation imperfections, provided that these imperfections do not exceed a certain threshold. This resilience is a promising aspect of implementing TQC in practical scenarios, where imperfections are often inevitable.

The critical perturbation thresholds for stability in TCs are significantly higher, by an order of magnitude, than those observed in TQCs, indicating that TCs exhibit markedly greater robustness to perturbations. This enhanced robustness is attributed to the phenomenon of quantum frequency locking~\cite{r27}, which occurs when a quantum system with an inherent characteristic frequency $\omega_c$ is driven at a frequency $\Omega = 2\pi/T$ that is close to a rational multiple $q/r$ of $\omega_c$. Under these conditions, the system can achieve stable oscillations with a period precisely aligned to an integer multiple of the driving period, $qT$.

In contrast, frequency locking in quasiperiodically driven systems does not occur in the simple and regular manner observed under periodic driving. Instead, more intricate forms of synchronization can arise when the driving frequencies are incommensurate (i.e., their ratios are irrational numbers). An analogous effect, sometimes referred to as irrational mode locking, has been reported in systems with a quasiperiodic spatial background \cite{pre23}. While the physical setting there differs from the temporally quasiperiodic driving considered in our study, our results demonstrate that a related mechanism can manifest in time, leading to richer frequency spectra. In particular, quasiperiodic driving allows the system’s response to lock onto linear combinations of the driving frequencies with integer coefficients but irrational overall ratios. This produces quasi stable oscillatory behavior that is structured yet aperiodic, reflecting locking onto these irrational frequency combinations.

\subsection{The prethermal TQC lifetime: effects of different system-sizes}

\begin{figure}
	\includegraphics[scale=0.24]{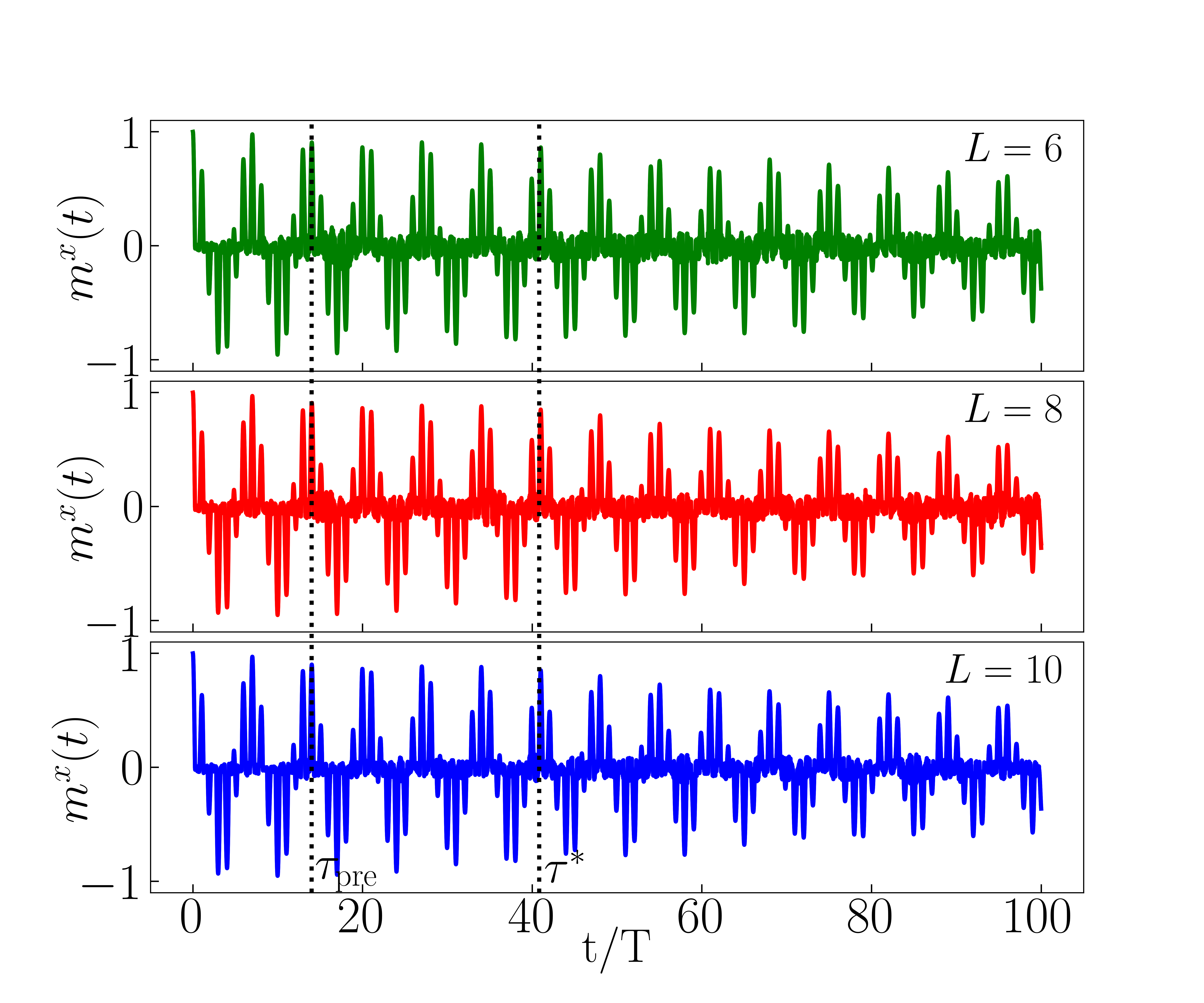}
	\centering
	\caption{The magnetization $m^x(t)$ for different system sizes at $\omega=1$ and $\Omega=\sqrt{2}/4$. The data show that the prethermal TQC lifetime, defined as $t_{\mathrm{th}}=\tau^{*}-\tau_{\mathrm{pre}}$, is approximately independent of $L$ within our numerical accuracy.}
	\label{Fig:qtc_diff-size}
\end{figure}

To examine the effect of system size on the lifetime of the prethermal TQC in the ITF chain, we compare the time evolution of the magnetization for chains of length $L = 6, 8$, and 10, as illustrated in Fig.~\ref{Fig:qtc_diff-size}. The results show that the TQC lifetime is approximately independent of the system size over this range.

As shown in Ref.~\cite{Kemp2017}, the edge spin in finite-size, open-boundary integrable spin chains—such as the ITF chain and the spin-$1/2$ $XYZ$ chain—can preserve the memory of its initial state over remarkably long times. Specifically, the expectation value of $\sigma^z_1$, representing the edge spin, stabilizes after an initial transient into a plateau that defines the edge spin coherence time. This coherence time, which originates from strong zero modes, grows exponentially with system size and depends sensitively on parameters like the exchange interaction $J$ and the transverse field $h$.

In our quasiperiodically driven model, the magnetization envelope also forms a plateau after an initial transient, characterizing the prethermal TQC lifetime. This plateau plays a role analogous to the edge coherence time described in~\cite{Kemp2017}. However, unlike the integrable case, the duration of this plateau does not show a pronounced dependence on system size, at least for the sizes $L = 6, 8, 10$ we studied. Instead, its lifetime is primarily controlled by drive parameters, such as the structure of the quasiperiodic modulation (see Fig.~\ref{Fig:qtc_diff-omega}), which dominate the stability of the prethermal TQC regime.

\subsection{The prethermal TQC lifetime: effects of derive frequencies}

In this section, we systematically examine how the prethermal time $t_{th}$ depends on the drive frequencies, by varying $\omega$ while maintaining a constant ratio $\Omega/\omega = \sqrt{2}/4$. 
This approach allows us to isolate the influence of an overall frequency scale without modifying the relative structure of the quasiperiodic driving.

\begin{figure}
	\includegraphics[scale=0.23]{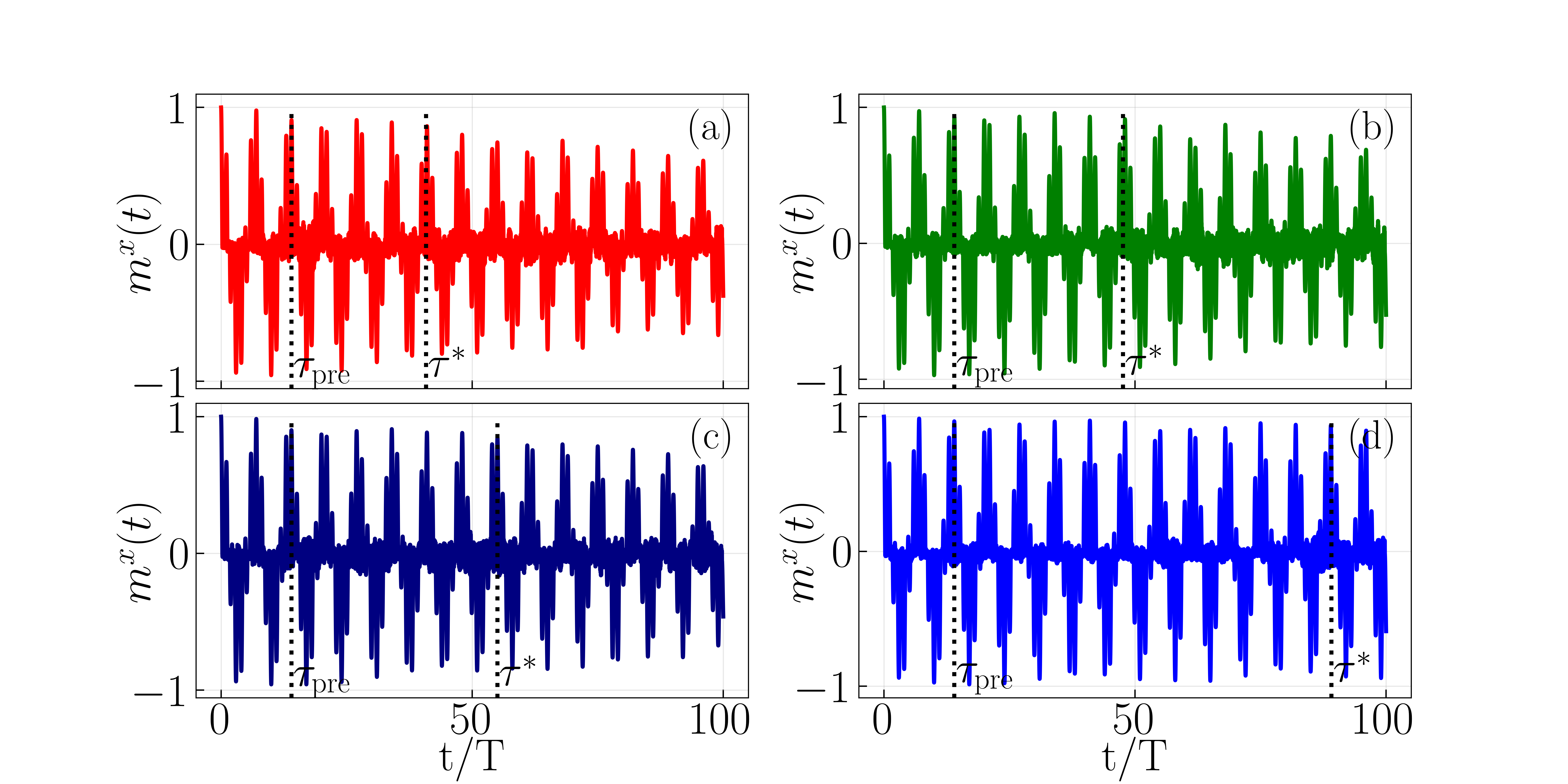}
	\centering
	\caption{Time evolution of the magnetization $m^{x}(t)$ for a quasiperiodically driven ITF chain of length $L=6$, with exchange couplings $J_i$ sampled from $[J/2, 3J/2]$ ($J=5.5$), and a transverse field $h=0.3$. The ratio is fixed at $\Omega/\omega=\sqrt{2}/4$. Panels (a) to (d) correspond to driving frequencies of $\omega = 1.0$, $1.25$, $1.5$, and $1.75$, respectively. The TQC lifetime increases with the driving frequency $\omega$, indicating that higher drive frequencies suppress heating, thereby enhancing the prethermal TQC lifetime.}
	\label{Fig:qtc_diff-omega}
\end{figure}

As shown in Fig. \ref{Fig:qtc_diff-omega}, increasing $\omega$ extends the plateau region between $\tau_{\mathrm{pre}}$ and $\tau^*$, which prolongs the prethermal TQC lifetime by delaying its thermalization.

The extension of the prethermal TQC lifetime is primarily due to the suppression of energy absorption processes in the high-frequency ($\omega$) limit, a characteristic feature of prethermalization in driven many-body systems. Specifically, increasing the driving frequency diminishes the probability of resonant transitions that could otherwise lead to rapid energy delocalization and heating. As a result, the system remains confined within a dynamically constrained manifold for an extended period, exhibiting robust quasiperiodic oscillations despite the absence of exact periodicity in the driving.

The extension of the prethermal TQC lifetime is more pronounced when $\omega \gg J$, where $J$ is the system's characteristic energy scale. This underscores the importance of a clear separation of timescales $\omega$ and $J$ ($\hbar=1$) for stabilizing the TQC phase. Our findings align with established theoretical arguments for prethermal behavior under high-frequency driving, applicable to both periodically and quasiperiodically driven systems\cite{pre1,pre2,pre3,pre4}. 

The current results underscore the crucial influence of the driving frequencies in determining the lifetime of the prethermal TQC. These findings not only reaffirm the prethermal character of the observed TQC dynamics but also provide a practical approach to enhance their experimental detectability through tuning the drive frequency.

\section{TQC phase in the ITF model with symmetric sampling of $J_i$s}\label{Sec:TQC-SG}

In this section, we examine the emergence of the TQC phase within the ITF model outlined in Eq. (\ref{Eq:ITF-Hamiltonian-quasi}), where the exchange couplings $J_i$ are randomly chosen from the symmetric interval $[-J/2, J/2]$.
This symmetric sampling implies that the exchange interactions can take both positive and negative values, leading to a rich landscape of magnetic behavior. 

For weak magnetic field strengths ($h \ll J$), the randomness in the signs of $J_i$ plays a crucial role in determining the magnetic properties of the system. Specifically, it has been shown that both the ground state and all excited states exhibit zero net magnetization \cite{r20}. This phenomenon arises because the competing interactions from the random couplings effectively cancel each other's contributions to the overall magnetization. 

Despite the absence of net magnetization, the system is not in a trivial paramagnetic phase; instead, it resides in a spin glass (SG) phase \cite{r24,r25,r26}. This conclusion is supported by the observation that the Edwards-Anderson order parameter is non-zero \cite{EA}, indicating the presence of localized magnetic order. 

In a spin glass phase, the system can sustain a finite density of domain walls, which become immobilized due to the underlying disorder. This immobilization results in locally broken symmetry, as different regions of the system may adopt distinct magnetic configurations \cite{r23,D.Huse}.
As a result, the eigenstates of the Hamiltonian are characterized by random glassy configurations, showcasing a complex interplay between disorder and magnetic interactions. These configurations reflect the frustration and variability inherent in the system, leading to a diverse array of possible states that can be accessed under different conditions. This behavior is characteristic of spin glasses, where the combination of randomness and interaction leads to a non-trivial energy landscape, challenging conventional notions of order in magnetic systems.

Consider a chain of length $L$ featuring the emergence of two domain walls located at sites $j_1$ and $j_2$. The quantum state corresponding to this configuration can be expressed in one of the following forms:
\begin{equation}
\nonumber|j_1, j_2\rangle=|\sigma_1,\dots,\sigma_{j_1-1},j_1,\sigma_{j_1+1},\dots,\sigma_{j_2-1},j_2,\sigma_{j_2+1},\dots,\sigma_L\rangle,
\end{equation} 
\begin{equation}
\nonumber|\bar{j}_1, \bar{j}_2\rangle=|\bar{\sigma}_1,\dots,\bar{\sigma}_{j_1-1},j_1,\bar{\sigma}_{j_1+1},\dots\bar{\sigma}_{j_2-1},j_2,\bar{\sigma}_{j_2+1},\dots,\bar{\sigma}_L\rangle,
\end{equation} 
where, $\sigma_i$ can randomly take values of $+1$ or $-1$, with the constraint that $\bar{\sigma}=-\sigma$. 
The above states are the eigenstate of the Ising Hamiltonian $H=\sum_iJ_i\sigma_i\sigma_{i+1}$, with random $J_i$. 
Under the influence of magnetic fields, the system experiences tunneling between the states $|j_1, j_2\rangle$ and $|\bar{j}_1, \bar{j}_2\rangle$. As a result, the Hamiltonian's eigenstate emerges as a coherent superposition of these states as:
\begin{equation}
|\{j_1,j_2\}, p=\pm1\rangle =\frac{1}{\sqrt{2}}(|j_1, j_2\rangle\pm|\bar{j}_1,\bar{j}_2\rangle).
\end{equation}
These states, characterized by the expectation values of the parity operator $\hat{P}$ and the Ising-symmetric domain wall operators $\hat{D}_{j_1} \equiv \hat{\sigma}^z_{j_1} \hat{\sigma}^z_{j_1+1}$ and $\hat{D}_{j_2} \equiv \hat{\sigma}^z_{j_2} \hat{\sigma}^z_{j_2+1}$, exhibit Ising symmetry breaking and long-range order, as evidenced by the non-zero spin-spin correlators in each eigenstate:
\begin{equation}
C_{ij}=\langle\{j_1,j_2\}, p|\sigma_i^z\sigma_j^z|\{j_1,j_2\}, p\rangle,~~\lim_{|i-j|\rightarrow a.l.}C_{ij}\neq0,
\end{equation}
where '$a.l.$' denotes arbitrarily large distances.

The driven disordered ITF model exhibits rich phases, particularly when compared to its undriven counterpart. In the undriven disordered ITF model, the primary phases observed are the PM and SG phases. However, the introduction of driving field leads to the emergence of additional dynamical phases.

In the ITF model periodically driven by an external magnetic field as in Eq. (\ref{Eq:itf-Hamiltonian}), the two notable dynamical phases are \cite{r19,r7}:

1. $\pi$PM-MBL phase: This phase is characterized by a paramagnetic behavior under the influence of a periodic driving field, where the system exhibits localization properties typical of MBL phases. In this regime, the quantum coherence and entanglement are preserved, despite the presence of disorder and external driving.

2. $\pi$SG-MBL phase: In this phase, the system displays spin glass characteristics while still retaining MBL traits. The dynamics are influenced by the interplay between the disorder, the transverse field, and the driving, leading to a scenario where the system can remain localized while exhibiting complex spin configurations typical of spin glass states.

In the periodically driven spin glass phase, our findings demonstrate a TC phase despite interaction perturbations and imperfect rotations. The system persists in the localized phase for interaction strengths ranging from $0<\lambda\ll 1$ \cite{r7}, highlighting its resilience to these perturbations while maintaining its position in the TC phase.

Shifting focus to the quasiperiodic spin chain governed by the Hamiltonian in Eq. (\ref{Eq:ITF-Hamiltonian-quasi}), we examine the time-dependent Hamiltonian $\hat{H}_{ITF}(t)$ as expressed in Eq. (\ref{Eq:ITF-Hamiltonian-MBL}). The couplings $J_i$ are drawn from the range $[-J/2, J/2]$, and the external fields $h_i$ are selected from $[0, h]$. The effects of the external driving field, represented by the Hamiltonian $\hat{H}_{drive}(t)$, are also incorporated into the analysis.

\begin{figure}
	\includegraphics[scale=0.27]{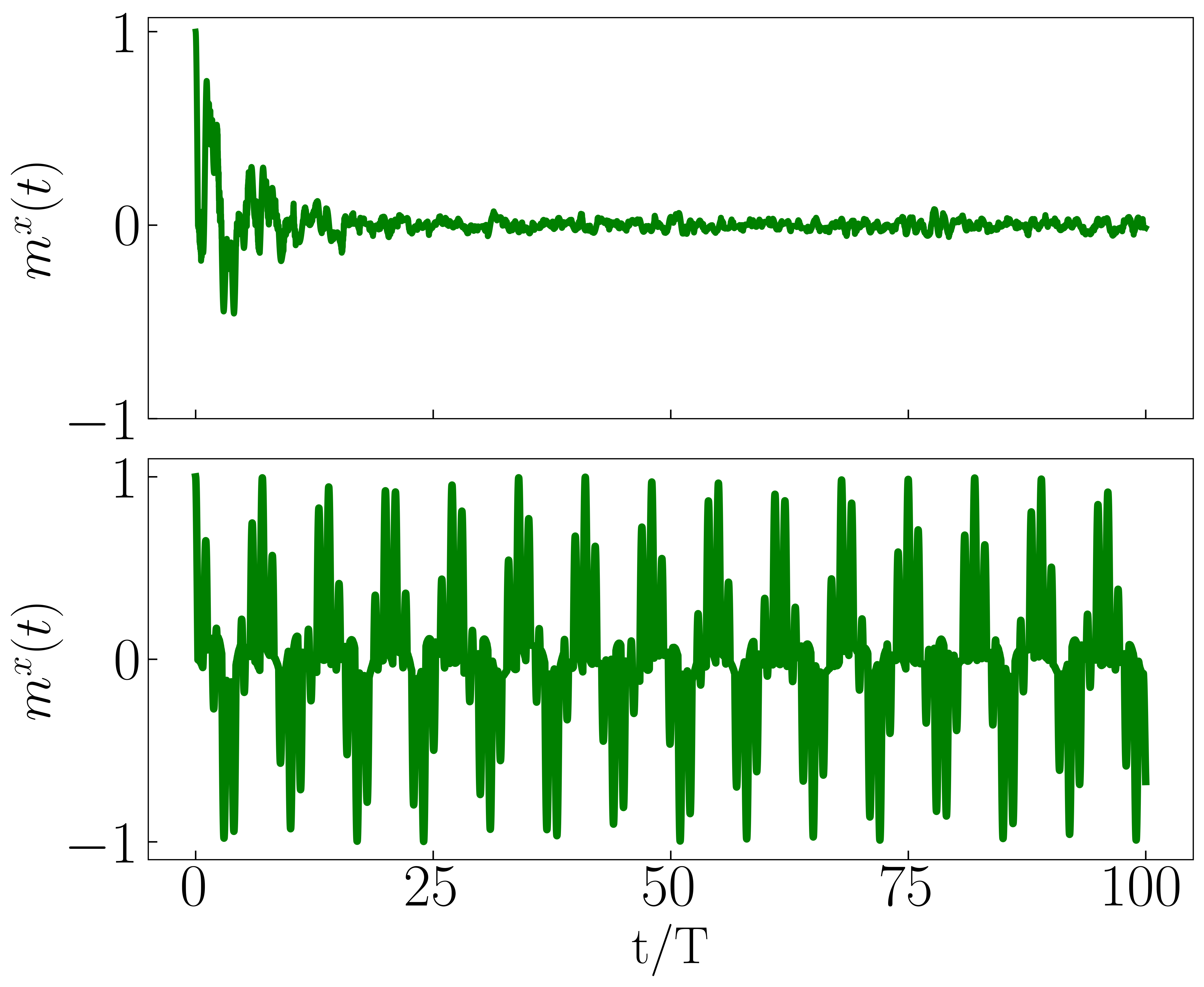}
	\centering
	\caption{Time evolution of the magnetization $m^{x}(t)$ for a quasiperiodically driven ITF chain of length $L=8$. The exchange couplings $J_i$ are sampled symmetrically from $[-J/2, J/2]$, with $J=5.5$, and the transverse field is set to $h=0.3$. (Up) $\omega=1$; (down) $\omega=12$, with $\Omega/\omega=\sqrt{2}/4$ in both cases. }
	\label{Fig:SG}
\end{figure}

The time evolution of the magnetization, as depicted in Fig. \ref{Fig:SG} (up), reveals a notable absence of TQC in either the SG or the PM phases under the condition of $\omega=1$ and $\Omega=\sqrt{2}/4$. This observation is in stark contrast to scenarios where $J_i$ is selected from the range $[J/2, 3J/2]$, where quasiperiodic dynamics can manifest. 

The lack of a TQC in the present setup can be primarily attributed to the intrinsic properties of the spin glass phase. Specifically, when $\delta J$ is drawn from a symmetric distribution, fluctuations in the couplings equally promote and inhibit local interactions, enhancing energy delocalization and accelerating thermalization. 

Interestingly, we find that increasing the driving frequencies $\omega$ and $\Omega$, while maintaining a fixed ratio $\Omega/\omega=\sqrt{2}/4$, leads to more pronounced and longer-lived quasiperiodic magnetization oscillations (see Fig.\ref{Fig:SG}-down). This observation highlights that both the choice of the coupling distribution $J_i$ and the values of the driving frequencies play a crucial role in stabilizing the TQC phase. 

The dynamical robustness of the TQC phase is highly sensitive to the drive structure and the underlying disorder. A detailed understanding of the stability boundaries and the influence of drive parameters, on the lifetime of the prethermal TQC remains an open and promising avenue for future research.

\section{Conclusion and outlook}\label{Sec:Conclud}

In summary, our investigation provides a detailed understanding of the dynamical response of a disordered quantum Ising chain in a transverse magnetic field (ITF) under both periodic and quasiperiodic driving.
In the periodically driven scenario, we confirmed that the system exhibits a robust discrete time crystal (TC) phase, characterized by persistent period-doubled oscillations in the magnetization $m^z(t)$. These oscillations remain stable despite the presence of interaction perturbations and imperfections in the driving protocol. Additionally, we demonstrated that increasing the chain length further enhances the stability of the TC phase.

In the quasiperiodically driven case, we observe the emergence of complex, structured oscillations at multiple incommensurate frequencies. These oscillations persist for a long time but eventually decay, indicating that the TQC phase corresponds to a transient prethermal dynamical regime rather than a stable thermodynamic phase.

The prethermal TQC lifetime is determined by the duration of the first plateau in the magnetization envelope. Our finite-size analysis reveals that this lifetime is largely independent of system size, at least for the chain lengths studied, demonstrating the robustness of the prethermal quasiperiodic order against finite-size effects.

Furthermore, the prethermal TQC lifetime exhibits a strong dependence on the quasiperiodic drive frequency. Our results show that increasing the drive frequency significantly extends the TQC lifetime.

Moreover, we found that TQC phase is suppressed when the exchange couplings are sampled symmetrically around zero, pointing to the destructive role of domain walls and emphasizing the importance of disorder structure. Notably, the quasiperiodic response remains stable under interaction perturbations and rotation imperfections, indicating the robustness of the observed prethermal regime over experimentally relevant timescales. Interestingly, we also observe that increasing the drive frequencies can lead to the reemergence of magnetization oscillations, suggesting that the drive frequencies play a pivotal role in enhancing the lifetime of the TQC phase. This behavior highlights the interplay between the drive parameters and disorder in shaping the dynamical stability of TQCs.

Our results enrich the current understanding of TC behavior in interacting spin systems and suggest clear signatures for experimental realization in platforms such as cold atomic gases and quantum simulators. The dynamical interplay between disorder, driving, and coherence revealed in our work opens avenues for studying novel non-equilibrium phenomena.

An intriguing future direction, inspired by time-quasiperiodic topological superconductors~\cite{pre18}, is to investigate whether TQC phases in spin chains can host robust edge modes or pairing operators. While such features have been predicted in fermionic systems via mechanisms like Majorana multiplexing, their analogs in spin systems with quasiperiodic driving remain an exciting open question for future exploration.

While a systematic determination of heating rates under general quasiperiodic drives remains an open problem, our results suggest that for the specific driving protocol considered here, such a characterization could be feasible and is an interesting direction for future work, particularly to better understand the persistence and decay of quasiperiodic oscillations in relation to energy absorption. In particular, identifying the precise boundaries of drive parameters that critically influence the lifetime of the prethermal TQC remains an open question.

Quasiperiodicity combined with topology can give rise to robust edge modes; notably, Majorana modes, topologically protected excitations, are central to realizing exotic and stable quantum phases \cite{pre18,pre19}. In the context of the disordered ITF chain studied here, we have not identified definitive evidence of well-defined edge states or pairing operators associated with the TQC phase. Nonetheless, the question of whether dynamically protected edge states can exist in quasiperiodically driven interacting spin systems remains highly intriguing.

For instance, in the paper \cite{pre19}, time-quasiperiodic topological superconductors were investigated, demonstrating that a system driven at $d$ mutually irrational frequencies can host up to $2^{d}$ types of Majorana modes. These modes can coexist despite spatial overlap and the absence of time-translational invariance. The authors introduced the concept of Majorana multiplexing, whereby quasiperiodic driving induces multiple Majorana edge modes at different quasienergies related to the incommensurate drive frequencies. This indicates that quasiperiodicity does not necessarily hinder the formation of topologically protected edge structures.

Moreover, it has been shown that Majorana edge modes, both zero-energy and those at golden-ratio quasienergy, can also appear in spin systems, such as the Fibonacci-driven quantum Ising chain \cite{pre18}. 

However, extending these concepts to disordered, many-body interacting spin systems under quasiperiodic driving with multiple irrational frequencies remains an open challenge. In particular, whether an analogue of Majorana multiplexing or other topological edge phenomena can manifest in many-body localized or prethermal spin chains driven quasiperiodically is still unknown. This question is particularly relevant to our model, where a disordered interacting Ising chain is subjected to a quasiperiodic drive involving multiple incommensurate frequencies.


\section{Acknowledgment}

We would like to express our gratitude to Seyed Saeed Soyouf Jahormi for his insightful discussions. Additionally, JA extends his appreciation to the Institute for Advanced Studies in Basic Sciences for their financial support through research Grant No. GIASBS12969.

\end{document}